# Analysing the impact of satellite constellations and ESO's role in supporting the astronomy community

Andrew Williams[1]
Olivier Hainaut[1]
Angel Otarola[1]
Gie Han Tan[1]
Giuliana Rotola[1]

[1] ESO

In the coming decade, up to 100 000 satellites in large constellations could be launched into low Earth orbit. The satellites will introduce a variety of negative impacts on astronomy observatories and science, which vary from negligible to very disruptive depending on the type of instrument, the position of the science target, and the nature of the constellation. Since the launch of the first batch of SpaceX's Starlink constellation in 2019, the astronomy community has made substantial efforts to analyse the problem and to engage with satellite operators and government agencies. This article presents a short summary of the simulations of impacts on ESO's optical and infrared facilities and ALMA, as well as the conducted observational campaigns to assess the brightness of satellites. It also discusses several activities to identify policy solutions at the international and national level.

## Introduction

The optical astronomy community was taken largely by surprise upon the launch of the first batch of SpaceX's Starlink satellite constellation on 23 May 2019. In their immediate post-launch configuration the 60 satellite units were visible as a very bright "string of pearls" travelling at high velocity across the night sky, and generated substantial public, media and astronomy community interest. ESO, along with many other observatories, national agencies, societies and the International Astronomical Union (IAU) issued public statements[1]. The community quickly became aware of the plans of many other companies and nations to develop similar constellations.

A satellite constellation is defined as "a number of similar satellites, of a similar type and function, designed to be in similar, complementary, orbits for a shared purpose, under shared control" (Wood, 2003). Currently operating constellations serve a variety of important and crucial functions for society, including: navigation and geodesy (for example, GPS, Galileo and GLONASS), satellite telephony (for example, Iridium), internet and TV (for example, ViaSat, Orbcom, GlobalStar) and Earth observation (for example, Copernicus and Planet). In the future, companies such as SpaceX, Amazon, Samsung, Telesat and OneWeb, and several national entities (for example, the Chinese and Indian Space Agencies) are planning very large constellations in low Earth orbit (LEO). These systems aim to provide low-latency broadband internet around the world to support the "Internet of Things" to connect directly machines and systems, financial and gaming transactions, and military applications. Their ultimate goal is to provide global high-bandwidth connectivity, including to remote places such as the middle of an ocean or a remote village (Curzi, Modenini & Tortora, 2020).

We were asked by the ESO Director General to analyse the impacts on ESO's facilities and to support the emergent community efforts with national societies and the IAU to study the issue and identify mitigations. In this article, we present a short summary of the outcome of the simulations of the impact on ESO's optical and infrared facilities and the observational campaigns conducted, and discuss several activities to develop policy solutions at the international and national level.

## Impacts on ESO facilities and astronomical science

### Visible and infrared spectral range

Drawing on public filings to the International Telecommunications Union (ITU) and also national regulatory agencies, we estimated the number of planned satellite constellations and the numbers of their individual units, as shown in Table 1. The following work uses the Starlink 1st and 2nd generations, OneWeb and GuoWang as a representative worst case scenario, totalling over 60 000 satellites between altitudes of 300 and 1200 kilometres. We used three independent methods to evaluate the number of satellite trails that will cross a field of view, as a function of the instrument characteristics, the pointing azimuth and altitude, the local time of night, and the time of year. The methods can either produce statistics for the whole or numerical values for specific directions of observation, fields of view and integration times. The three methods give good agreement; further technical details are available in Walker et al. (2020). Figure 1 displays the result of one of these simulations. The effect on observations scales with the number of satellites, so the effect of the current ~ 2000 constellation satellites is about 3.5% of what is discussed below.

| Constellation (Registering Nation) | Altitude (km) | Number of satellites |
|---|---|---|
| Starlink Generation 1 updated (US) | 550 | 1584 |
|  | 540 | 1584 |
|  | 570 | 720 |
|  | 560 | 348 |
|  | 560 | 172 |
| Starlink Generation 1 Phase 2 (US) | 335 | 2493 |
|  | 341 | 2478 |
|  | 346 | 2547 |
| Starlink Generation 2 (US) | 328 | 7178 |
|  | 334 | 7178 |
|  | 345 | 7178 |
|  | 360 | 2000 |
|  | 373 | 1998 |
|  | 499 | 4000 |
|  | 604 | 144 |
|  | 614 | 324 |
| OneWeb Phase 2 reduced (US. UK) | 1200 | 1764 |
|  | 1200 | 2304 |
|  | 1200 | 2304 |
| Amazon Kuiper (US) | 590 | 784 |
|  | 610 | 1296 |
|  | 630 | 1156 |
| Guo Wang GW-A49 (China) | 590 | 480 |
|  | 600 | 2000 |
|  | 508 | 3600 |
|  | 1145 | 1728 |
|  | 1145 | 1728 |
|  | 1145 | 1728 |
|  | 1145 | 1728 |
| Sat Revolution (Poland) | 350 | 1024 |
| CASC Hongyan (China) | 1100 | 320 |
| CASIC Xingyun Lucky Star (China) | 1000 | 156 |
| CommSat (China) | 600 | 800 |
| Xinwei (China) | 600 | 32 |
| AstromeTech (India) | 1400 | 600 |
| Boeing (US) | 1030 | 2956 |
| LeoSat (Luxembourg) | 1423 | 108 |
| Samsung (Korea) | 2000 | 4700 |
| Yaliny (Russia) | 600 | 135 |
| Telesat LEO (Canada) | 1000 | 117 |
| Total |  | 78 265 |

Table 1. Planned satellite constellations. This table reflects publicly available filings for spectrum from the International Telecommunication Union (ITU) and national communications regulators. Projects are at varying stages of approval. The data for Starlink and OneWeb include recent changes filed at the US Federal Communication Commission (FCC). Some of the operators have withdrawn their applications (for example, Boeing, LeoSat). Only Starlink and OneWeb have launched operational satellites (over 1737 and 218, respectively, as of June 2021). Many more companies have filed applications for other purposes such as remote sensing. As these are typically much smaller — and hence fainter — than telecommunication satellites, they are not considered in the analysis.

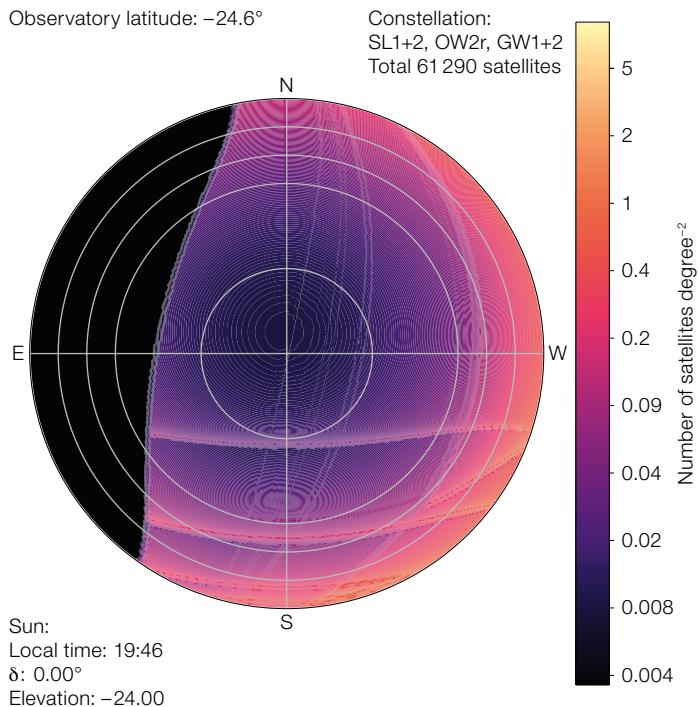

Figure 1. (Upper left) An example distribution of satellites over Paranal observatory. This figure shows a map of the sky above Paranal at the beginning of an equinox night, showing a representative distribution of Starlink, OneWeb, and GuoWang satellites. Of the total of 61 290 satellites, 2562 are above the horizon, including 1373 illuminated ones. The colour indicates the density of illuminated satellites (in objects degree$^{-2}$). The east-west features correspond to the edges of constellations whose inclinations are close to the latitude of the site. The dark crescent to the East marks the area where all satellites are already in the Earth's shadow.

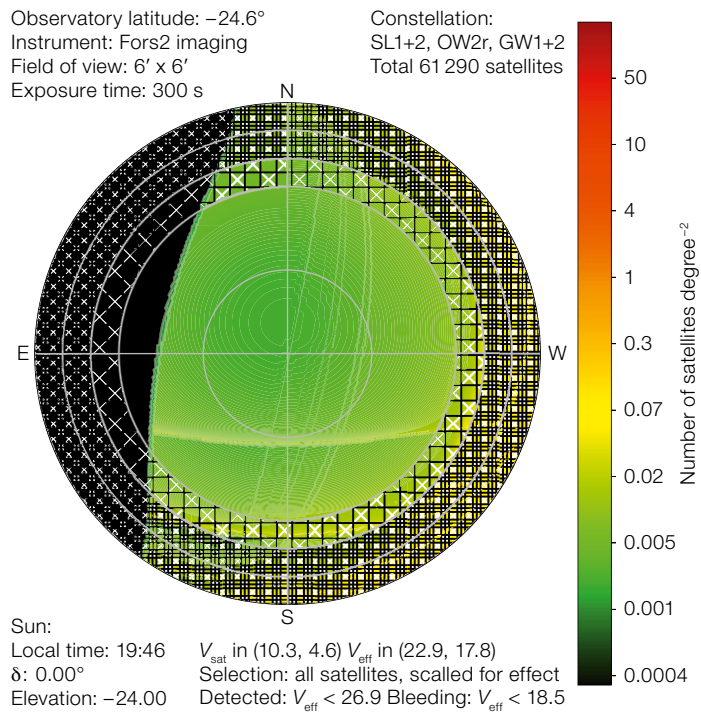

Figure 2. (Upper right) The effect of the satellites on an exposure. As in Figure 1, this map of the sky above Paranal observatory shows the effect of the satellites on a 300-second image obtained with the FOcal Reducer and low dispersion Spectrograph (FORS), as a function of the direction of pointing. The high airmass regions are hatched. The colour indicates the fraction of exposures lost.

To evaluate the losses to an observation, we counted the number of satellite trails crossing the field of view during the exposure, considering the brightness and apparent angular velocity of each satellite. We excluded those that are too faint to be detected. Those that are detected destroy a 5-arcsecond-wide trail across the observation, and those saturating the detector destroy the whole observation. Figure 2 shows an example for a 300 second image with FORS. The losses, averaged over zenithal distances below 60 degrees (airmass 2 and better, representing 95% of observations with the Very Large Telescope), are presented in Figure 3 for a series of representative instruments. Because of the apparent concentration of satellites towards the horizon, the values in the figure should be about doubled when setting the limit at a zenithal distance of 70 degrees, and halved when considering only 30 degrees around the zenith. The figure shows that most instruments will suffer losses at the ≲ 1% level at twilight, that figure dropping by 1–2 orders of magnitude when the sun reaches an elevation of –40 degrees. Because of their brighter limiting magnitudes, the high-resolution spectrographs are essentially immune to the satellites considered. The low-resolution spectrograph will be able to register some of the satellites as low-signal-to-noise contamination, which could be difficult to disentangle from the science signal. Vera C. Rubin Observatory, with its 3 degree imager behind an 8-metre telescope, is particularly affected. Its situation is made worse by the effect of saturated trails on the camera, resulting in losses of up to about 30% for observations at twilight.

The situation in the thermal infrared is completely different: the satellites are emitters, so they remain a source of contamination even when in Earth's shadow, and they are well above the detection threshold of instruments like the VLT Imager and Spectrometer for mid-InfraRed (VISIR) (Hainaut & Williams, 2020). However, the field of view of a thermal IR instrument is small, and the individual exposure time is very short, so only a negligible fraction of the exposures will be affected. A satellite passing in front of a star will cause a short eclipse. However, because of their high apparent angular velocity, it lasts of the order of 1 millisecond, and will cause

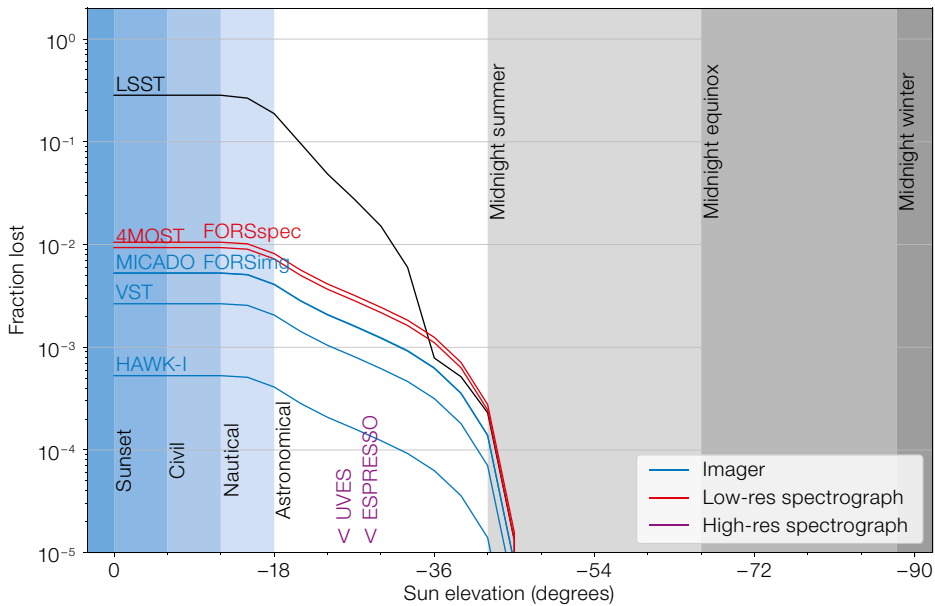

Figure 3. The mean fraction of exposure lost due to satellite trails as a function of the solar elevation, for representative exposures obtained at zenithal distances smaller than 60 degrees. The twilights and the solar elevation at midnight are indicated with shading. The exposures are as follows: the Vera C. Rubin Observatory LSST Camera — 15 seconds; the 4 metre Multi-Object Spectrograph Telescope (4MOST) on the Visible and Infrared Survey Telescope for Astronomy (VISTA) — fraction of contaminated fibres for 1200 seconds; the VLT Survey Telescope (VST) — 300 seconds; FORS — 1200 seconds for spectroscopy, 300 seconds for imaging; the High Acuity Wide-field $K$-band Imager (HAWK-I) — 60 seconds; the Multi-AO Imaging Camera for Deep Observations (MICADO) — 60 seconds; the Ultraviolet and Visual Echelle Spectrograph (UVES) — 1 hour, the Echelle SPectrograph for Rocky Exoplanet and Stable Spectroscopic Observations (ESPRESSO) — 1 hour. The simulations account for the limiting magnitude and saturation magnitude of the systems; they assume that a detected satellite ruins a 5-arcsecond-wide trail across the detector, and that a saturated satellite ruins the whole exposure. In the case of fibre-fed spectrographs (ESPRESSO, 4MOST), a detected satellite ruins the exposure. For 4MOST, a multi-fibre spectrograph, the number of trails in the field is converted into the number of affected fibres using a Monte-Carlo simulation which indicates that on average a satellite will affect 1.254 fibres. FORS images are affected at a level comparable to that of the much wider field VST because it is sensitive to much fainter satellites. In the cases of ESPRESSO and UVES, the spectrographs are not sensitive to the fast-moving satellites.

a detectable dip in the signal of the star only for very fast photometry (< 1 second).

Several mitigations have been identified that can reduce the impacts on astronomical facilities. The first option would be to reduce the number of satellites, lower their orbital altitude and minimise their size. Existing constellation designs, however, are already highly optimised and linked to the spectrum assigned by regulatory authorities. Post-hoc changes are likely to be challenging, and therefore the community should focus on influencing operators in the early design stages. Other operator mitigations focus on design measures to darken satellites and avoid reflected sunlight. Observatories can introduce scheduling based on the simulation results (for example, pointing to the darkest areas in maps like that of Figure 2), or consider shutter-control measures and additional small telescopes, or suitable wide-field cameras, to detect satellites during or just before observations so as to either avoid them or identify contaminated data. The level and complexity of the telescope or operations mitigations put in place must be commensurate to the effect they correct. For many types of observations, it might be cheaper and more efficient just to repeat a failed observation. Currently, the impact on ESO visible and infrared facilities will remain low enough that no telescope/instrument or scheduling mitigations are foreseen, but we are closely monitoring developments.

The astronomy community, particularly in the US, has had a very productive collaboration with SpaceX aimed at reducing the brightness of satellites. SpaceX trialled a dark coating on "DarkSat" Starlink, which achieved some brightness reduction, but had implications for the thermal control and was subsequently abandoned in favour of attitude adjustments and a new "VisorSat[2]". This features the addition of a sunshade which protects the body of the satellite from direct illumination, and which is mounted so that the illuminated side of the shade is not visible from Earth. Preliminary measurements suggest darkening by a factor of about 3 compared to first-generation satellites. VisorSats have been launched since mid-2020. Adjustments of the attitude of the satellites have also reduced their brightness: the orientation of the solar panel has been adjusted to keep it invisible from the ground, hidden behind the bus of the satellite.

### Millimetre and submillimetre spectral range

Of the satellite constellations listed in Table 1 the Starlink Generation 1 may impact ALMA as downlink transmitters since the end-user terminals are planned to operate at frequencies between 37.5 and 42.5 GHz, which is within the ALMA Band 1 observing range. Whilst operator agreements and observation scheduling could avoid the unlikely case of direct illumination of the ALMA antennas, the cumulative background noise will be a more persistent problem. We assume that the interference from each satellite adds in an uncorrelated way in an ALMA receiver, meaning that the signal power of each satellite can just be added to obtain the total power received from all satellites. Under these assumptions a total noise increase due to the Starlink Generation 1 satellites of about 50 milli-Kelvins at the input of an ALMA Band 1 receiver is expected. Longer integration times should compensate for this noise. Where this cannot be done, reductions in sensitivity will have to be accepted, although this requires more detailed study. If a significant noise increase is expected, this should be included in the ALMA Sensitivity Calculator so that the correct integration times can be derived. Further details are available in a report made for ESO's 155th Council Meeting[3].

The Starlink Generation 2 constellation does not mention the use of a downlink to user terminals in ALMA Band 1; this function is foreseen at lower frequencies where ALMA does not observe. However, a request has been made as part of the US Federal Communication Commission (FCC) licence application to operate a gateway downlink in the frequency range 71–76 GHz which falls in ALMA Band 2. Since the number of gateways will be much less than the number of end-user terminals, interference to ALMA is expected to be a lesser concern. This assumes that these gateway terminals are not located close to the ALMA observatory. Planning the location of the gateway terminals needs licensing from the Chilean national authorities and it is recommended that the Joint ALMA Observatory be actively involved in this process. At the time of writing, SpaceX has submitted a request to the Chilean regulatory body, SubTel, to operate four gateways on Chilean territory.

### Satellite observations

Satellite observations have provided an important complement to simulations to calibrate assumptions and to characterise brightness over a range of wavelengths and from different locations and orbital geometries. Satellites exhibit highly dynamic changes in apparent brightness that are due to geometrical and operational factors, requiring many observations to capture the changes in brightness over the full range of parameters. Assessing impacts on astronomy also requires knowledge of satellites' apparent velocity, position in the sky, and frequency of sightings.

Observations of the brightness magnitude of the Starlink and OneWeb satellites have been made for the purposes of a) determining how bright they are in various astronomical spectral bands of interest, and b) in the case of the Starlink satellites, determining how effective are the engineering design changes implemented to decrease their brightness and meet the recommendations of the SATCON1 (Walker et al., 2020) and the Dark and Quiet Skies for Science and Society (IAU 2020) workshops. SpaceX launched a mitigation satellite, Starlink-1130[a] (dubbed DarkSat), that included a special darkening treatment of the Earth-facing sides of the satellite structure. This was intended to make it less reflective of sunlight and effectively dimmer.

The satellites Starlink 1130 (DarkSat) and Starlink 1113[b], launched on 7 January 2020, were observed on 3 March 2020 immediately after arriving at their nominal operational orbital height of 550 kilometres. The analysis of the observations produced a brightness magnitude (in Sloan Digital Sky Survey $g'$ filter, scaled to zenith and range 550 kilometres) of $5.33 \pm 0.05$ and $6.10 \pm 0.04$ magnitudes for Starlink 1113 and 1130 (DarkSat), respectively (Tregloan-Reed et al., 2020). These results showed that a) the reduction in brightness for DarkSat was $50.8\% \pm 3.5\%$, b) the brightness magnitude of DarkSat (6.10 magnitudes) was still within the range of the naked-eye limiting magnitude of an experienced observer (Bortle, 2001) and c) DarkSat was still brighter than the limiting magnitude of 7 that has been recommended to minimise unwanted effects in sensitive astronomical cameras (Walker et al., 2020; IAU 2020). Subsequent observations of these satellites were conducted in various spectral bands to assess their brightness and the effectiveness of the darkening treatment used in DarkSat, in various astronomical spectral bands from visible to near-infrared (Tregloan-Reed et al., 2021). The results of these observations, conducted using the Chakana telescope (Char et al., 2016) and VIRCAM (Emerson, McPherson & Sutherland, 2006; Dalton et al., 2006) on the VISTA telescope, are summarised in Figure 4. The results shown in Figure 4 illustrate that the non-darkened Starlink satellites are brighter, and that the effectiveness of the darkening treatment decreases with increasing wavelength.

For completeness, observations of the Starlink VisorSat (IAU 2020) measure a magnitude in the visible similar to that reported here for DarkSat. Observations of OneWeb satellites (Tregloan-Reed et al., in preparation), operating at an orbital height of 1200 kilometres, show that their brightness in the visible is about 8 magnitude. However, in 46% of these observations the satellites are brighter than the SATCON1 recommendation of 7.85 magnitude for a satellite at this altitude.

### Policy activities and the way forward

The simulations and observations outlined here have been facilitated by the coordinated efforts of several national societies, such as the American Astronomical Society, the Royal Astronomical Society and the European Astronomical Society, amongst others, along with input from observatories such as ESO, Vera C. Rubin Observatory, NSF's NOIRLab and the Square Kilometre Array Observatory. This body of work has established a basis for understanding the problem space and yielded promising early results on operator mitigations. The bilateral work with companies is welcome but is not a sustainable solution as the number of private and public constellation projects around the world grows. Astronomers are now beginning to look for regulatory solutions at both international and national levels.

At the international level, ESO supported the IAU in submitting a set of policy recommendations[4], developed in the Dark and Quiet Skies project (IAU 2020), to the United Nations Committee on the Peaceful Uses of Outer Space (COPUOS), which addresses international dialogue on governance of the use and exploration of outer space. The recommendations call for, *inter alia*, international agreement to address impacts on astronomy by minimising the orbital altitude, the brightness of space objects to less than unaided-eye levels (7 magnitude), and antenna sidelobe emissions such that their indirect illumination of radio observatories and radio-quiet zones does not interfere, individually or in aggregate. Most importantly, the work aimed to create a norm of cooperation within the astronomy community and consideration of impacts on science and on the dark sky at the design and early regulatory approval of projects. The paper, which was co-signed by the IAU and five countries (Chile, Ethiopia, Jordan, Slovakia and Spain), was submitted to the 58th session of the Science and Technical Sub-Committee of COPUOS held on 19–30 April 2021. Many COPUOS members voiced support for the astronomy community's concern and the need to find solutions, and the Committee agreed that the IAU should continue to study the matter further and report back to the 2022 session.

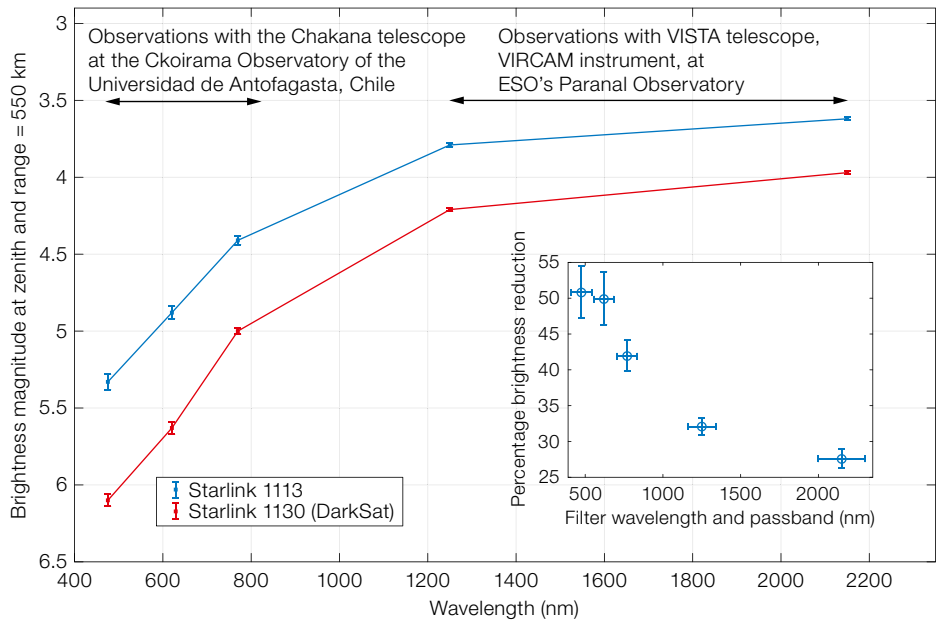

Figure 4. Brightness magnitude of Starlink satellites 1130 (DarkSat) and 1113. Data extracted from Tregloan-Reed et al., (2020, 2021). The insert shows the computed reduction in brightness for Starlink-1130 (DarkSat), similarly to Figure 6 in Tregloan-Reed et al. (2021).

Whilst COPUOS is an important forum in which to raise awareness and possibly agree non-binding guidelines, a binding international treaty to implement the IAU policy guidelines is highly unlikely. Instead, the astronomy community must look to national regulatory authorities. In this regard, the primary issue is that, apart from a narrowly-focused law in the US preventing "space billboards[5]" no states have yet regulated optical pollution from space, and the ITU does not include optical frequencies in its Radio Regulations. A concern for radio astronomers is that so-called Radio-Quiet Zones, while offering protection from ground-based radio emitters, do not cover space-based radiation sources. Several activities to look for regulatory solutions are proceeding, starting with the US-led SATCON2 project[6] to explore the possibilities for legal protection under environmental law, in addition to addressing national launch and communications regulators and standards bodies.

Despite the partial success of recent operator mitigations and some possibilities for introducing regulations to compel operators to coordinate with astronomers and make design changes to satellites, no combination of mitigations can fully avoid impacts on astronomy. With the possibility of 100 000 satellites launching in the coming decades, the impacts on astronomy are one concern amongst many in this new megaconstellation era (Boley & Byers, 2021). This large number of satellites in LEO creates a major concern in terms of orbital crowding, collision avoidance and control of debris. Whilst various space agencies are considering the problem, the existing space governance system is being stretched to its limits. We hope the astronomy community, along with all space actors and beneficiaries of pristine dark and quiet skies, will work towards a shared stewardship of the night sky in a way that supports conservation, economic development, science and exploration and sustainability of the environment.

### References

Bortle, J. E. 2001, Sky & Telescope, February 2001
Boley, A. C. & Byers, M. 2021, Sci Rep, 11, 10642
Char, F. et al. 2016, BAAA, 58, 200
Curzi, G., Modenini, D. & Tortora, P. 2020, Aerospace, 7(9), 133
Dalton, G. B. et al. 2006, Proc. SPIE, 6269, 62690X
Emerson, J., McPherson, A. & Sutherland, W. 2006, The Messenger, 126, 41
Hainaut O. R. & Williams, A. P. 2020, A&A, 636, A121
IAU 2020, *Dark and Quiet Skies for Science and Society — Report and Recommendations*
Tregloan-Reed, J. et al. 2020, A&A, 637, L1
Tregloan-Reed, J. et al. 2021, A&A, 647, A54
Walker, C. et al. 2020, Bull. AAS, 52(2)
Wood, L. 2003, *Satellite constellation networks, Internetworking and Computing over Satellite Networks*, Boston: Springer, 13

### Links

[1] ESO public statement: https://www.eso.org/public/announcements/ann19029/
[2] SpaceX's VisorSat: https://spacenews.com/spacex-to-test-starlink-sun-visor-to-reduce-brightness/
[3] Report presented to the ESO Council: http://www.eso.org/public/about-eso/committees/cou/cou-155th/external/Cou_1928_Satellite_Constellations_161120.pdf
[4] Policy recommendations submitted to COPUOS: https://www.iau.org/static/publications/uncopuos-stsc-crp-8jan2021.pdf
[5] US law on space advertising: https://www.law.cornell.edu/uscode/text/51/50911
[6] Announcement of SATCON2 workshop: https://aas.org/posts/news/2021/05/satellite-constellations-2-workshop-announced

### Notes

[a] Starlink 1130 satellite, NORAD ID 44932, COSPAR ID 2020-001U.
[b] Starlink 1113 satellite, NORAD ID 44926, COSPAR ID 2020-001N.